\documentclass[aip,reprint]{revtex4-2}

\usepackage{graphicx}
\usepackage{xfrac}
\usepackage{graphics,epsfig,graphicx}

\usepackage{easyreview}
	\usepackage{todonotes}
	\usepackage{siunitx}
	\usepackage{physics}
	\usepackage{epstopdf}
	\usepackage{verbatim}
	\usepackage{amsmath}
	\usepackage[unicode=true, bookmarks=true, bookmarksnumbered=false, bookmarksopen=false, breaklinks=false, pdfborder={0 0 1}, backref=false, colorlinks=true]{hyperref}
\usepackage{cleveref}
	
	\usepackage{color} 

\usepackage[normalem]{ulem} 
\usepackage{color} 

\begin{document}
\title{Microwave cavity-free hole burning spectroscopy of Er$^{3+}$:Y$_2$SiO$_5$ at millikelvin temperatures}

\author{A.~Mladenov}
\affiliation{Experimentalphysik, Universität des Saarlandes, D-66123 Saarbrücken, Germany}
\affiliation{Institute for Nuclear Research and Nuclear Energy, Bulgarian Academy of Sciences, Sofia}

\author{N.~Pankratova}
\affiliation{Department of Physics, Joint Quantum Institute and Center for Nanophysics and Advanced Materials, University of Maryland,  USA}

\author{D.~Sholokhov}
\affiliation{Experimentalphysik, Universität des Saarlandes, D-66123 Saarbrücken, Germany}



\author{V.~Manucharyan}
\affiliation{Department of Physics, Joint Quantum Institute and Center for Nanophysics and Advanced Materials, University of Maryland, USA}

\author{R.~Gross}
\affiliation{Experimentalphysik, Universität des Saarlandes, D-66123 Saarbrücken, Germany}
\affiliation{Walther-Meißner-Institut, Bavarian academy of sciences, Garching, Germany}
\affiliation{Munich Center for Quantum Science and Technology (MCQST), Munich, Germany}
\affiliation{Technical University of Munich (TUM), Munich, Germany}

\author{P.~A.~Bushev}
\affiliation{Experimentalphysik, Universität des Saarlandes, D-66123 Saarbrücken, Germany}
\affiliation{JARA-Institute for Quantum Information (PGI-11), Forschungszentrum Jülich, 52428 Jülich, Germany}

\author{N.~Kukharchyk}
\affiliation{Experimentalphysik, Universität des Saarlandes, D-66123 Saarbrücken, Germany}
\affiliation{Walther-Meißner-Institut, Bavarian academy of sciences, Garching, Germany}
\affiliation{Munich Center for Quantum Science and Technology (MCQST), Munich, Germany}

\email[E-mail: ]{nadezhda.kukharchyk@wmi.badw.de}
\date{\today}

\begin{abstract}
{\bf
	Efficient quantum memory is of paramount importance for long-distance  
	quantum communications, as well as for complex large-scale computing  
	architectures. We investigate the capability of Er$^{3+}$:Y$_2$SiO$_5$ crystal to serve as a quantum memory for the travelling microwave photons by employing techniques developed for dense optical ensembles. In our efforts to do so, we have  
	performed high-resolution microwave spectroscopy of Er$^{3+}$:Y$_2$SiO$_5$, where we  
	identified electronic spin as well as hyperfine transitions.  
	Furthermore, we have explored spectral hole burning technique and studied the spin relaxation process at millikelvin temperatures, determined the main relaxation mechanisms, which lay the groundwork for further studies of the topic.
}
\end{abstract}
\maketitle

\section{Introduction}
\label{sec:intro}
Microwave regime in quantum technologies is currently best represented by the superconducting circuits based on superconducting qubits~\cite{Krantz.2019}. While the quantum processing units based on such superconducting qubits demonstrate outstanding results~\cite{Arute.2019}, further enhancement of calculation capacities addresses the necessity of effective microwave quantum memory~\cite{Gouzien.2021}. 
At the same time, the longest storage times are demonstrated with optical quantum memories\cite{Zhong2015,Zhong2017}, which, however, rely on ultra-long coherence of the spin states of the rare earth spin ensembles.

Realization of purely microwave quantum memories has been demonstrated with electronic spins ensembles based on donors in silicon and rare earth ions~\cite{Ranjan.2021}. Depending on the choice of the quantization axis and magnetic field working point, storage times in such ensembles may reach six hours~\cite{Zhong2015}. Such extremely long storage time is possible due to using the zero first-order Zeeman (ZEFOZ) transitions, which are analogous to the clock transitions and possess strongly reduced sensitivity to the fluctuations of the external decoherence sources.
Coupling of the microwave fields to the spin ensembles in microwave experiments is typically performed via resonating structures~\cite{Probst2015,Probst2013,Tkalcec.2014,Weichselbaumer2020}, which limits the number of transitions accessible at the same time.  

Ability to effectively control the electronic spins with the propagating microwave signals~\cite{Calajo.2016,Jen.2021,Hsu.2016,Saglamyurek.2011,Fedorov2021}, which can couple to the same spins in broad frequency range, will pave the way to a large variety of microwave quantum memory schemes including more than two electronic levels. 
The atomic frequency comb (AFC) protocol is one of the most promising quantum memory protocols established in optics\cite{Etesse.2021,Simon.2010,Teja.2021}, which can be realized on two electronic levels as well as extended to a third longer living state also allowing for frequency domain multiplexing storage protocols~\cite{Afzelius2009,Sinclair2014}.
Deterministic narrow spectral hole burning (SHB) is a necessary step in realization of AFC. 
While these techniques are strongly established in optics, they have not yet been widely studied in the microwave domain. The first attempt in burning microwave spectral holes has been performed on spin ensembles coupled to a resonator~\cite{Putz.2017}.

In this work, we explore the spectral hole burning with the propagating microwave signals in Er:Y$_2$SiO$_5$ crystal in the cavity-free regime, which allows for addressing the electronic spins in a broad frequency range. We characterize the control efficiency in our experimental scheme, identify the relaxation rates and processes, and extract the resulting spectral hole profiles. We address the potential of the SHB with the propagating microwave signals for the microwave quantum memory techniques.

\section{Experimental detail }
\begin{figure*}[t]
	\centering
	\includegraphics[width=0.7\textwidth]{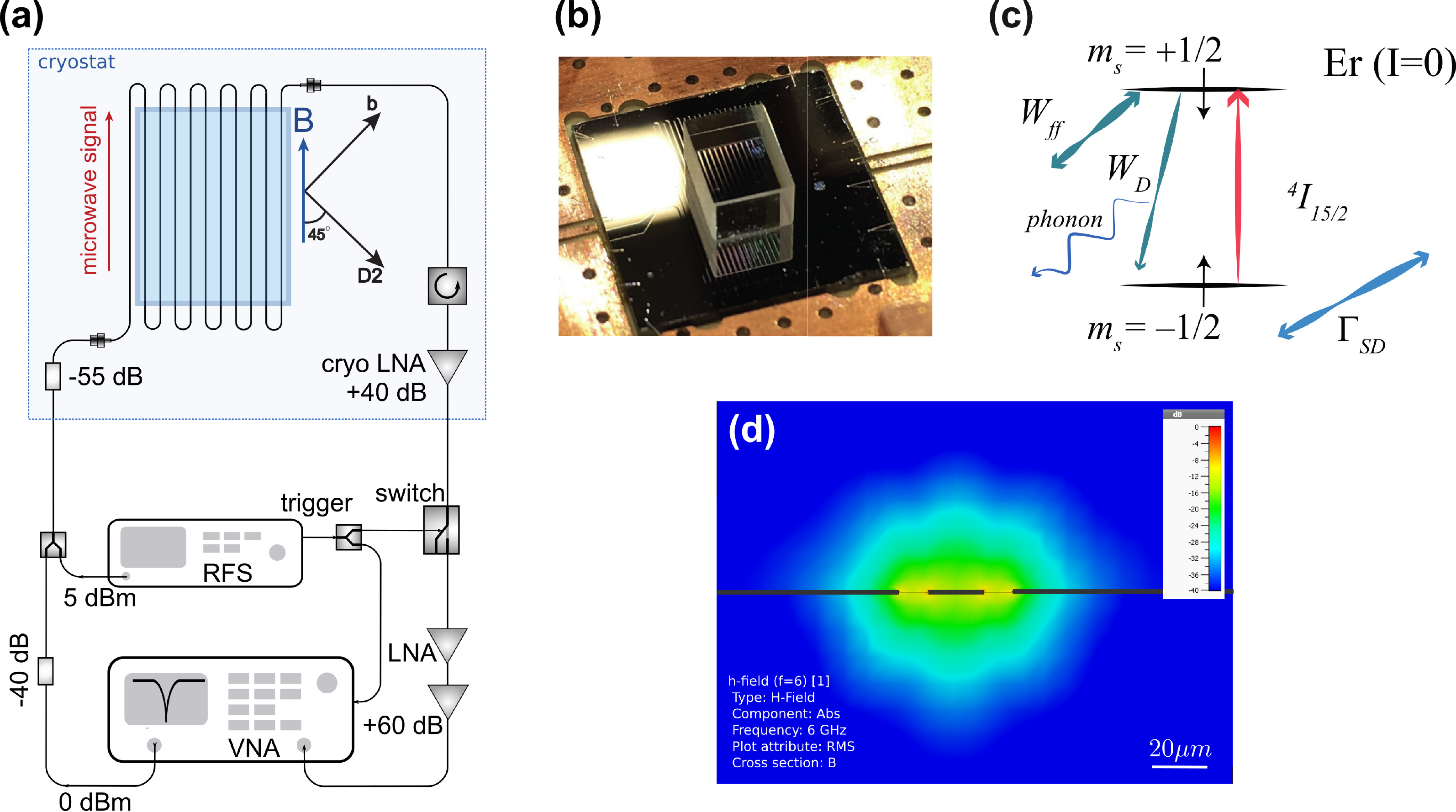}
	\caption{(Colour online) (a) Experimental scheme: Er:Y$_2$SiO$_5$ crystal is placed on top of the SC chip with meandered co-planar transmission line. The DC magnetic field is applied along the surface of the chip. 
		VNA is the radio-frequency vector network analyser, which supplies the probing signal for spectroscopy and hole recovery measurement. RFS is a radio frequency source, which supplies the hole burning pulse. The VNA and a microwave switch are triggered by the RFS. 
		(b) Photo of the crystal on the meandered transmission line. (c) Illustration of the level scheme and the relaxation processes responsible for the relaxation. $W_{ff}$, $W_D$ and $\Gamma_{SD}$ stand for the flip-flop process, direct process and spectral diffusion, respectively. (d) Simulation of the microwave mode propagating in a coplanar transmission line used in experiment, which is shown as an RMS value of the propagating magnetic field component. Simulation is performed in CST Studio Suite.}
	\label{fig:experiment}
\end{figure*}

We investigate the Erbium-doped Y$_2$SiO$_5$ (YSO) crystal, which is supplied by Scientific Materials Inc. with 0.02$\%$ of Erbium atomic concentration. Erbium is in its natural abundance with $\sim23\%$ being $^{167}$Er ($I=\sfrac{7}{2}$), and $\sim77\%$ being 5 isotopes with zero nuclear spin. The crystal has dimensions $\SI{3}{\milli\meter}\times\SI{4}{\milli\meter}\times\SI{5}{\milli\meter}$ and is placed on the top of the co-planar superconducting transmission line, see Fig.~\ref{fig:experiment}(a,b). The Erbium spins are inductively coupled to the transmission line via magnetic field created by the propagating microwave signal.
The orientation of the optical crystal axes is shown in Fig.~\ref{fig:experiment}(a). The external magnetic field vector and axes D2~\cite{Wyon1992} and b are located in the plane of the SC chip with an angle of 45$^\circ$ between axis b and magnetic field vector. The magnetic field vector is parallel/anti-parallel to the wave-vector of the microwave field. Due to the low crystalline and point symmetry of YSO crystal, Erbium ions occupy two crystallographic nonequivalent sites, each showing two magnetically nonequivalent transitions. The orientation of the crystalline axes with respect to the magnetic field vector has been selected to fully lift the degeneracy between the magnetically nonequivalent sites.

Schematics of the measurement scheme is shown in Fig.~\ref{fig:experiment}(c). In the microwave spectroscopy and in hole burning measurements, the probing signal is created by Vector Network Analyzer (VNA) at the power of 0\,dBm, which is additionally attenuated by \SI{-95}{\decibel} before reaching the sample. After the crystal, the transmitted signal is amplified by cryogenic low noise amplifier by \SI{+40}{\decibel} and by room-temperature low noise amplifiers by \SI{+60}{\decibel}. 

Microwave hole-burning pulse is generated by a radio-frequency source (RFS) in the power range of 5\,dBm to 15\,dBm  with a time duration duration of \SI{0.5}{\second} to \SI{5}{\second}. 
The burning pulse is attenuated on the input RF lines of the cryostat by \SI{-55}{\decibel}. 
At the end of the burning pulse, VNA starts measuring time dependence of the transmission at a selected frequency. In order to protect the VNA input, it is cloaked from the high-power burning pulses by an RF switch, with both VNA and switch being triggered by the RFS. 
Both, change of absorption amplitude and phase are recorded as functions of time with intermediate frequency filter bandwidth (IFBW) of \SI{10}{\hertz} and sampling interval of $\sim$\SI{50}{\milli\second}.

\section{Microwave spectroscopy}
\begin{figure*}[t]
	\centering
	\includegraphics[width=0.9\textwidth]{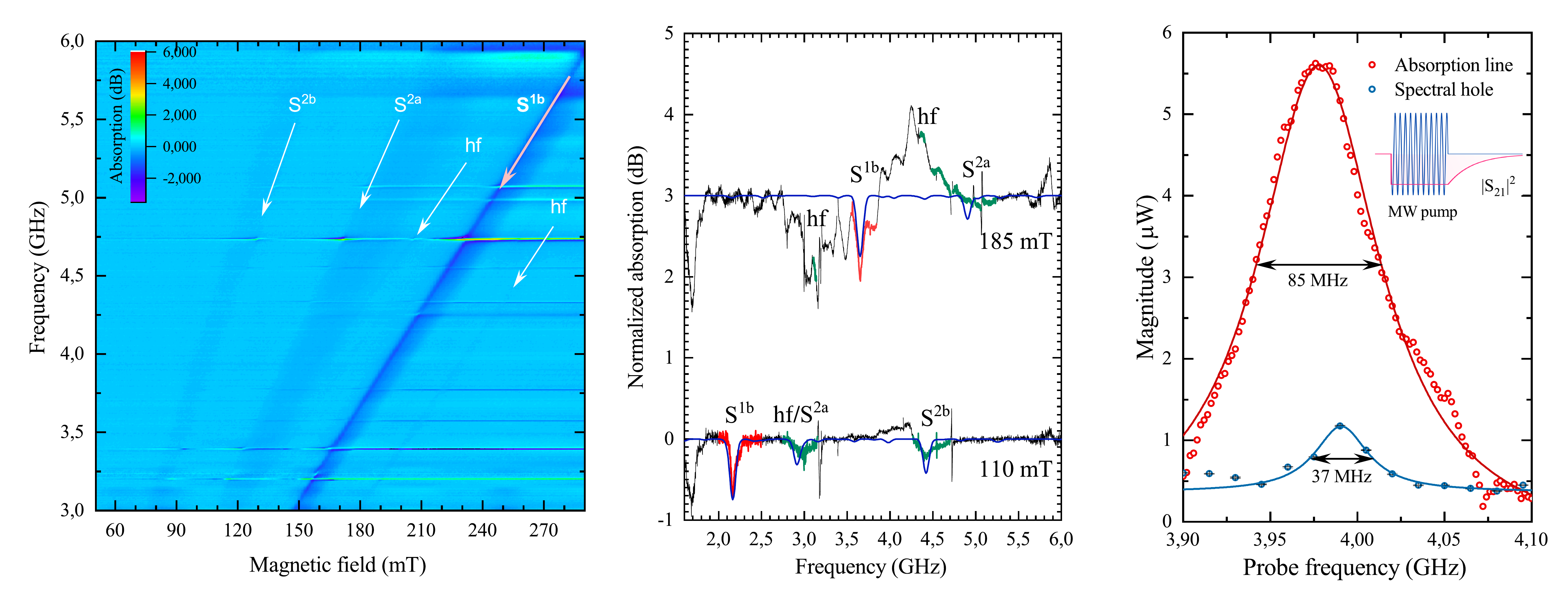}
	\caption{(Colour online) (a) 3D plot of the microwave absorption spectrum on magnetic field and probing frequency. The absorption lines are marked according to the identified microwave transitions. Hole burning spectroscopy is performed on the strongest S\textsuperscript{1b} transition. Additional resonances at fixed frequencies are due to reflections on cable connections and bonding on the sample holder and the chip with the transmission line, see Fig.~\ref{fig:experiment} for the measurement setup details.\\
		(b) Measured (black lines) and simulated (blue lines) absorption spectra at selected magnetic fields of \SI{110}{\milli\tesla} and \SI{185}{\milli\tesla}. The identified transitions are highlighted with light blue in the measured spectra. The S\textsuperscript{1b} transition studied with hole burning technique is highlighted with red color. Amplitude of the simulated spectra is given in arbitrary units.\\
		(c) Measured absorption profiles (circles) and Lorentzian fits (solid lines) of the absorption line and extracted spectral hole at \SI{167}{\milli\tesla}. The inset depicts measurement sequence for the spectral hole recovery.}
	\label{fig:spectra}
\end{figure*}

Microwave spectroscopy is performed on I$_{15/2}(0)\ket{-\frac{1}{2}}~\leftrightarrow~$I$_{15/2}(0)\ket{+\frac{1}{2}}$ spin transition as a function of magnetic field at a base temperature of cryostat equal to \SI{10}{\milli\kelvin}, see \cref{fig:spectra}(a). The absorption spectra are measured as S$_{21}$ parameter of VNA. In the spectrum, three strong absorption lines correspond to the crystallographic (S$^1$ or S$^2$) and magnetic (S$^{\_a}$ or S$^{\_b}$) nonequivalent sites of the erbium ions and are identified with  the available g-tensor~\cite{Sun2008_g-tensor} as marked in \Cref{fig:spectra}(a,b). The fourth absorption line, S\textsuperscript{1a}, possesses the largest g-factor of $\simeq13$ and is outside the measured frequency range at applied magnetic fields. The hyperfine transitions of $^{167}$Er isotope with the nuclear spin of $\sfrac{7}{2}$ are observed at a much smaller absorption amplitude. The most pronounced absorption line corresponds to the S\textsuperscript{1b} site, on which we investigate the hole-burning technique in this work.

The structure of microwave spectrum is determined by the Zeeman, hyperfine and quadrupole interactions, where the two last ones are only present for $^{167}$Er. The full spin hamiltonian then looks the following:
\begin{equation}
	\mathcal{H} = \mu_B \vb{B}\cdot\vb{g_s}\cdot\vb{S} + \vb{S}\cdot\vb{A}\cdot\vb{I} +\vb{I}\cdot\vb{Q}\cdot\vb{I}, 
\end{equation}
where $\mu_B$ is the Bohr magneton, vectors $\vb{S}$ and $\vb{I}$ are the electron and nuclear spin numbers, respectively. For the ground state of erbium, electron spin equals $S=\sfrac{1}{2}$, and nuclear spin of $^{167}$Er equals $I=\sfrac{7}{2}$. Tensors $\vb{g_s}$, $\vb{I}$ and $\vb{Q}$ represent matrix elements for electron Zeeman, hyperfine and quadrupole interactions, respectively.
To simulate the full spectrum, we take advantage of the EasySpin simulation package~\cite{Stoll2006_Easyspin}, where we specify the $\vb{g_s}$, $\vb{I}$ and $\vb{Q}$ tensors, taken from \cite{Sun2008_g-tensor}, and fraction of erbium isotopes with $\textrm{I}=0$ and  $\textrm{I}=\sfrac{7}{2}$ in the naturally abundant spin ensemble. 
When comparing the measured g-factors to the simulated values, we find misalignment angles, which are about $3^{\circ}$ in b-D2 plane and $0.5^{\circ}$ in D1-b plane for site 1, and $6^{\circ}$ for site 2 in both planes. Small misalignment in b-D2 plane can be partially attributed to a misalignment of the crystal on the SC chip. It cannot explain, however, large discrepancy on site 2. Similar misalignment angles have been observed by Welinski et al.~\cite{Welinski2016}, where the discrepancy on the site 2 has been attributed to a rotation of the C2 point-symmetry axis of the substituting rare earth ion with respect to the original Yttrium ion. 

Comparison of the measured microwave spectrum against the one, simulated with corrected axes, is shown in \cref{fig:spectra}(b) for magnetic fields of \SI{110}{\milli\tesla} and \SI{185}{\milli\tesla}, while numeric values for the extracted g-factors are given in \cref{tab:g-fact}. 

\begin{table}[h!]
	\centering
		\caption{Overview of g-factor values obtained from simulated g-tensor and from the experiment.}
		\label{tab:g-fact}
	\begin{tabular}{l| c | c | c | c |}
		& S\textsuperscript{1a} & S\textsuperscript{1b} & S\textsuperscript{2a} & S\textsuperscript{2b}  \\	
		\hline
		simulated~\cite{Sun2008_g-tensor} & 13.11 & 1.41 & 1.89 & 2.87 \\ 
		measured & - & 1.41 & 1.87 & 2.87 \\
	\end{tabular}

\end{table}

The further absorption spectrum analysis and hole-burning investigation are performed on the strongest transition S\textsuperscript{1b} with a g-factor of 1.41, which is composed of \SI{42}{\percent} of \textsuperscript{168}Er, \SI{35}{\percent} of \textsuperscript{166}Er, and \SI{23}{\percent} \textsuperscript{170}Er. To characterize the line, we convert it to the amplitude units and fit to the Lorentzian lineshape function, an exemplary fit is shown in \cref{fig:spectra}(c). 
The extracted linewidth values depend on magnetic field as $\Gamma_{\mathrm{FWHM}} = \Gamma_{0} + \overline{\delta\gamma} B$, where $\Gamma_{\mathrm{FWHM}}$ is the full width at half maximum, $\Gamma_{0}$ is the linewidth at zero magnetic field and is found to be \SI{17(4)}{\mega\hertz}, and $B$ is the magnetic field. The inhomogeneity of the gyromagnetic ratio, $\overline{\delta\gamma}$, equals to \SI[per-mode = symbol]{0.21}{\mega\hertz\per\milli\tesla}, which is \SI{1.1}{\percent} of the g-factor value. The $\Gamma_{\mathrm{FWHM}}$ is also in agreement with previous publications if taking into account the magnetic field working point~\cite{Probst2013,Probst2014}.
Temperature of a spin ensemble, often addressed as an effective temperature,  becomes an important parameter at millikelvin temperatures. Even small excitation result into elevated effective temperatures of spin ensembles, which can be an order of magnitude higher then the temperature measured on the cryostat, which makes it necessary to identify real effective temperature of the spin ensemble. The key issue with the spin temperature is that the spin levels are strongly quantized with respect to the thermal energy level of the surrounding thermal bath. As a result, the effective spin ensemble temperature is defined as a probability of finding the electronic spin in a particular energy level, and it is calculated via Boltzmann distribution for a canonical ensemble~\cite{waldram1985,Gemmer2010}.

To find the temperature of the spin system in the spectroscopy measurement, we fit the area under the absorption line to the Boltzmann distribution on temperature, given as $\sim \exp(\sfrac{\Delta E }{k_B T_s})/(1+\exp(\sfrac{\Delta B }{k_B T_s}))$, where $\Delta E$ is the energy difference between the spin states, $k_B$ is the Boltzmann constant, and $T_s$ = \SI{81.9(20)}{\milli\kelvin} is the spin temperature.
At the same time, the base temperature on the cryostat plate is measured by a sensor to be equal to \SI{11}{\milli\kelvin}. Thermal interface from the spin ensemble to the cryostat is mediated by spin-phonon interaction (phonon bath) and thermal boundary resistance (Kapitza resistance). The later one lead to a significant suppression of the thermal energy flow from spin ensemble to the cooling element and result into such a difference in temperature.

\begin{figure*}[th!]
	\centering
	\includegraphics[width=0.9\textwidth]{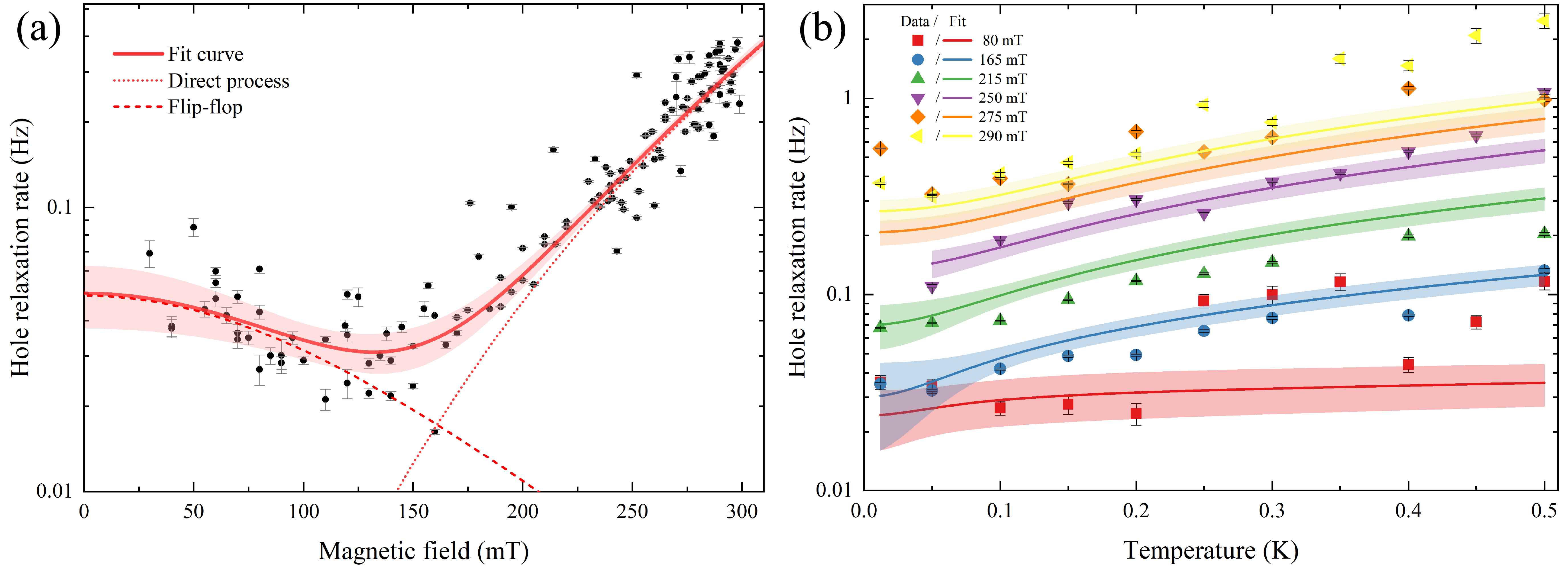}
	\caption{The spectral hole relaxation times as a function of magnetic field (a) and temperature (b). Dots and filled symbols represent the experimental data, the solid lines show the fits of experimental data to the model, and red dashed and dotted lines outline the contributions of the flip-flop and direct processes to the total relaxation rate.}
	\label{fig:field_temp}
\end{figure*}

\section{Dynamics of spectral holes}
Spectral hole burning is performed in a saturation recovery experiment, when a microwave burning pulse of a fixed frequency is sent to the crystal for a fixed time. The population of the ground state is then transferred to the excited state, as indicted with the red arrow in \cref{fig:experiment}(c). As a result, the absorption amplitude is reduced. The recovery of the saturation amplitude over time is detected with a week probing signal of VNA and is governed by the dynamics of relaxation processes.

\paragraph*{Relaxation processes.} 
Below \SI{1}{\kelvin}, spin relaxation is limited to two main processes: flip-flop, $R_{\textrm{FF}}$, and direct process, $R_{\textrm{d}} = \frac{1}{\tau_\textrm{d}}$.
These processes are schematically shown in \cref{fig:experiment}(c). Third process depicted in \cref{fig:experiment} is spectral diffusion, $\Gamma_{SD}$, which contributes into broadening of the spectral hole.

The spin-spin relaxation rate given by flip-flop process is defined as~\cite{Abragam1970_spin-spin}
\begin{eqnarray}
	R_{\textrm{FF}}=w_{\textrm{FF}}\tanh[2](\frac{g \mu_\textrm{B} B}{2 k_\textrm{B} T_{\textrm{s}}}),
\end{eqnarray}
where $w_{\textrm{FF}}$ is the flip-flop coefficient, which, following the approach from Car et al.~\cite{Thiery2019_flip-flop}, we estimate to be equal \SI{50}{\milli\hertz}. Constants $\mu_\textrm{B}$ and $k_\textrm{B}$ are Bohr magneton and Boltzmann constant, respectively, and $T_{\textrm{s}}$
is the effective temperature of the spin system. 

Rate of the direct process over the magnetic field and temperature in absence of the phonon bottleneck is given by~\cite{Abragam1970_spin-phonon}
\begin{eqnarray}
	R_{\textrm{d}} = \frac{1}{\tau_{\textrm{d}}} = w_\textrm{d}g^5\textrm{B}^5\coth(\frac{g \mu_\textrm{B} B}{2 k_\textrm{B} T_{\textrm{s}}}), 
\end{eqnarray}  
where the spin-phonon relaxation rate coefficient $w_\textrm{d}$ is estimated to be equal to $\SI{23}{\hertz\tesla^{-5}}$~\cite{Abragam1970_spin-phonon,Kukharchyk2019,Probst2015}. 

Thus, the final equation including flip-flop and direct process relaxation rates reads as $R = R_{\textrm{FF}}+	R_{\textrm{d}} $, which is then fit independently to the magnetic field, \cref{fig:field_temp}(a), and temperature, \cref{fig:field_temp}(b), dependent data. Derived coefficients are listed in \cref{tab:relax-par} and show very good agreement with analytically derived values.

\paragraph*{Effective temperature.} 
The effective temperature of the spin ensemble $T_s$, which is derived from relaxation processes, is in a good agreement with the spin temperature derived from the microwave spectrum,  $\sim \SI{70}{\milli\kelvin}$, see \cref{tab:relax-par} for detail. It also means that below magnetic fields of $B_{pol} \simeq \SI{2}{\frac{\tesla}{\kelvin}}\cdot T_s\simeq \SI{140}{\milli\tesla}$ the spin ensemble is not polarized to the ground state. This results into stronger flip-flop rate as well as into smaller hole amplitudes leading to smaller signal-to-noise ratio in the data. Below the magnetic field of \SI{70}{\milli\tesla}, splitting of the spin states is smaller than the thermal energy level, ${k_B T_s} \ge {g \mu_B B} $, and saturation recovery is not measurable. 

Apart from extracting temperature of the spin ensemble, we directly measure the temperature of the cryostat itself which reaches down to $\SI{11}{\milli\kelvin}$. 
Temperature of the cryostat remains unchanged during the sweep of the magnetic field and is controllably changed during the temperature-dependent measurement. For fitting the relaxation processes to experimental data, we introduce the correction of the actual temperature of the spin ensemble with respect to that of the cryostat~\cite{Kukharchyk2019,Kukharchyk2020}:
\begin{equation}
	T_s = T_{min}(1+(\sfrac{T}{T_{min}})^2)^{\sfrac{1}{2}},
\end{equation} 
where $T$ is the temperature of the cryostat and $T_{min}$ is minimal temperature attainable by the spin system. At the limit of $T<T_{min}$, $T_s = T_{min}$, while for $T>T_{min}$, $T_s = T$.
Relaxation rates extracted thus from the temperature dependent data are in good agreement with the magnetic field dependence. Largest declination of the data points from the fit-curves are observed for the lowest and higher fields, see \cref{fig:field_temp}(b). However, such a straggle of the data points is similar to that observed for the multiple measurements in the magnetic field dependence, \cref{fig:field_temp}(a).

\begin{table}[b!]
	\centering
	\caption{Coefficients of relaxation processes and effective temperature values}
	\begin{tabular}{l| S |  S| S | c |}
		& $w_{D}$ & $w_{ff}$ & $T_{s}$ & R-Sq. \\
		& \si{\hertz} & \si{\milli\hertz}&  \si{\milli\kelvin} & \% \\
		\hline
		Fit of magnetic field dependence   & 23(1) &  50(1) & 69(8) & 89\\
		Fit of temperature dependence   & 22(2) &  30(1) & 73(21) & 85\\
		\hline
		Analytically derived    & 23 & 51 &  &  \\
		\hline
		Spectroscopy    &  &  & 81.9 & 50\\
		\hline
	\end{tabular}
	\label{tab:relax-par}
\end{table}

\paragraph*{Rabi frequencies and acting microwave power.} 
For efficient control of the a spin ensemble over the transmission line, we need to know the effective acting power of the magnetic field component of the microwave signal and correct Rabi frequency. 
To identify the Rabi frequency, we vary the length of the burning pulse and measure the maximally achieved amplitude of the spectral hole. From the observed oscillations, we derive the Rabi frequency to be equal to $\simeq \SI{3.9}{\hertz}$. Taking into account the size of the propagating microwave mode, see \cref{fig:experiment}(d), we derive the RMS amplitude of the magnetic component of the microwave field $B_{AC} \simeq \SI{0.35}{\nano\tesla}$, or total acting microwave power $P_{act-R} \simeq \SI{45}{\femto\watt}~(-105\,\textrm{dBm})$.

From the spectroscopy measurement, we also derive additional losses on the sample and SC to be in the order of \SI{-40}{\decibel}. Simulating the propagation of the microwave mode as shown in \cref{fig:experiment}(d), we estimate the mean loss of amplitude in the main acting volume of the microwave mode to be in the order of \SI{-20}{\decibel}. Applying the total loss from the spectroscopy measurement, comprising of $\sim \SI{-115}{\decibel}$, see \cref{fig:experiment}, to the microwave power of $15\,\textrm{dBm}$ generated by the RFS, we obtain expected acting power of $P_{act-a}=-100\,\textrm{dBm}$ in the microwave mode. We thus see that the acting powers from Rabi frequency, $P_{act-R}$, and from analytical calculations, $P_{act-a}$, are in good agreement.
The further hole burning experiment is performed with $5\,\textrm{dBm}$ output power of RFS, with Rabi frequency of $\simeq \SI{1}{\hertz}$, at which we have not observed any pronounced oscillations when varying the pulse duration.

\section{Spectral hole profiles}
\begin{figure*}[t!]
	\centering
	\includegraphics[width=0.9\textwidth]{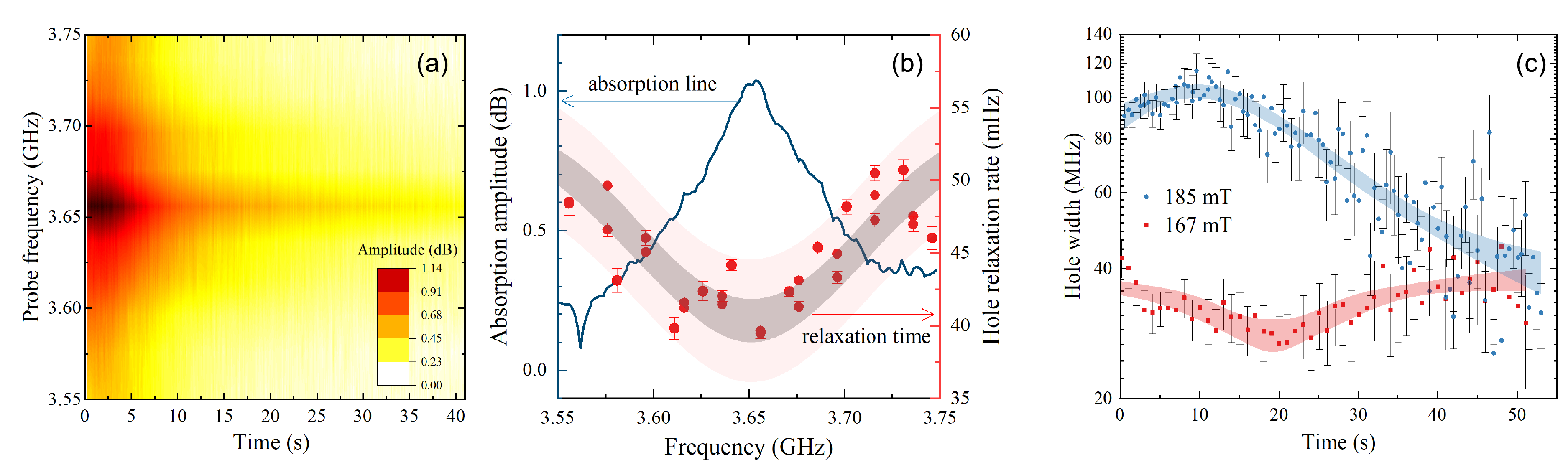}
	\caption{(a) Time-dependent profile of the microwave spectral hole a function of time and frequency detuning from the central burning frequency, and (b) relaxation time at different detuning frequencies as compared to the absorption line profile, which are measured at \SI{185}{\milli\tesla}.  
		(c) Spectral hole width as a function of relaxation time at \SI{167}{\milli\tesla} and \SI{185}{\milli\tesla}.
	}
	\label{fig:hole_profile_detail}
\end{figure*}

By keeping the burning frequency fixed and changing the probing frequency, we identify the profiles of the spectral holes. Such a three-dimensional spectrum of a spectral hole is shown in \cref{fig:hole_profile_detail}(a), where the pump frequency was kept fixed at \SI{3.651}{\giga\hertz} and probing frequency was scanned around the pump with total span of \SI{20}{\mega\hertz}.
When increasing the burning time, the amplitude of the spectral hole is increased, while the width of the spectral hole and relaxation time are independent of the burning time, which has been confirmed at several magnitudes of magnetic field. Therefore, the burning pulse length of \SI{3}{\second} has been selected for the hole burning experiment as an optimal one. Independence of the spectral hole width of the burning pulse length is attributed to the rate of spectral diffusion, which is faster ($\sim\SI{1}{\kilo\hertz}$) than possible shortest burning pulse (\SI{0.1}{\second}-\SI{1}{\second}). 

The profiles of spectral holes have been probed for a set of magnetic filed values. By fitting the profiles to the Lorentzian shape, we extract amplitude and width of the created spectral holes as functions of time, \cref{fig:hole_profile_detail}, and of magnetic field.
In the range of magnetic fields below \SI{170}{\milli\tesla}, area and amplitude of the holes are nearly constant. At increase of magnetic field above \SI{170}{\milli\tesla}, the width of spectral hole increases while amplitude decreases. The relative amplitude of spectral holes is at the level of $\sim\SI{7.5}{\percent}$ of the spectral line amplitude, and similarly the relative width slightly varies around $\sim\SI{65}{\percent}$ of the spectral line width. 

Measuring the relaxation of the spectral hole over its profile, we observed it to be faster when measured outside of width at half maximum for magnetic field of \SI{185}{\milli\tesla}, see \cref{fig:hole_profile_detail}(b). Inside of the spectral hole width region, the relaxation time can be considered constant. Such behavior of spectral holes appears then as narrowing of the spectral hole with time as shown in \cref{fig:hole_profile_detail}(c). Interestingly, the spectral hole burned at \SI{167}{\milli\tesla}, which profile is demonstrated in \cref{fig:spectra}(c), has shown nearly no change of the width over time. Moreover, there is a narrowing of the spectral hole after \SI{20}{\second} of relaxation, after which the width returns to it's initial value of $\simeq\SI{37}{\mega\hertz}$. The wider spectral hole observed at \SI{185}{\milli\tesla} converges to a width value of $\simeq\SI{40}{\mega\hertz}$, which is similar to that at \SI{167}{\milli\tesla} resulting approximately \SI{60}{\percent} of the absorption line width. At other magnetic fields, width of spectral holes is also $\sim 60\%$, revealing no change over the time. We suggest that the large initial broadening on spectral hole at \SI{185}{\milli\tesla} and it's subsequent narrowing is related to interference of one of the on-chip parasitic resonance at a close frequency ($\sim \SI{3.75}{\giga\hertz}$)  to the transition in-study. 

\section*{Conclusions}

In conclusion, we have for the first time demonstrated a cavity-free microwave spectral hole burning on Erbium-doped Y$_2$SiO$_5$ crystal. We were able to achieve efficiency of spectral hole of \SI{60}{\percent} in width and \SI{7.5}{\percent} in amplitude in the range of magnetic fields up to \SI{300}{\milli\tesla}. From the spectral hole recovery over time, main relaxation processes are identified: flip-flop process and direct process. The effective temperature of the spin ensemble from both spectral hole burning and spectroscopy converged to $\simeq\SI{70}{\milli\kelvin}$.
The achieved in this work results on spectral hole width and depth are limited by the duration of the hole-burning pulse and the acting microwave power. We have demonstrated the hole burning over a large span of magnetic fields and frequencies, accessible in absence of resonator. Further enhancement of the control of the spin systems in cavity-free regime require exploiting the zero first-order Zeeman transitions and increase of the acting microwave power in burning pulse. 
Results presented in this publication pave the way towards the realization of cavity-free microwave quantum memory based on rare-earth spin ensembles.

\section*{Acknowledgement}


\paragraph{Funding information}
This work is supported by DFG through the grants INST 256/415-1, BU 2510/2-1 and EXC 2111.

\bibliography{bibliography_SHB}

\end{document}